\documentclass[aps,prb,floatfix,twocolumn,showpacs,amssymb]{revtex4-1}
\usepackage{graphicx}
\usepackage{epsfig} 
\usepackage{amssymb}
\usepackage{siunitx}
\usepackage{bm}
\usepackage{color}
\newcommand{\be}{\begin{equation}}
\newcommand{\ee}{\end{equation}}
\newcommand{\bea}{\begin{eqnarray}}
\newcommand{\eea}{\end{eqnarray}}
\newcommand{\nn} {\nonumber}

\newcommand{\Tr}{ {\rm Tr} \, }
\newcommand{\vvr} { {\bf r} }

\def\G{{\rm \Gamma}}
\def\d{\delta}
\def\D{{\rm \Delta}}

\def\ve{\varepsilon}

\def\l{\lambda}
\def\L{{\rm \Lambda}}

\def\S{{\rm \Sigma}}

\def\vf{\varphi}
\def\F{{\rm \Phi}}

\def\w{\omega}

\def\bra{\langle}
\def\ket{\rangle}

\def\xc{{\rm xc}}
\def\x{{\rm x}}

\def\Tr{{\rm Tr}\,}

\begin{document}
\title{Local vertex corrections from exchange-correlation kernels with a discontinuity}
\author{Maria Hellgren}
\affiliation{Sorbonne Universit\'e, Mus\'eum National d'Histoire Naturelle, UMR CNRS 7590, IRD, Institut de Min\'eralogie, de Physique des Mat\'eriaux et de Cosmochimie, IMPMC, 4 place Jussieu, 75005 Paris, France}
\date{\today} 

\begin{abstract}
The fundamental gap of an interacting many-electron system is given 
by the sum of the single-particle Kohn-Sham gap and the derivative discontinuity. 
The latter can be generated by advanced approximations to the exchange-correlation 
(XC) energy and is the key quantity to capture strong correlation with density 
functional theory (DFT). In this work we derive an expression for the derivative 
discontinuity in terms of the XC kernel of time-dependent density 
functional theory and demonstrate the crucial role of a discontinuity 
in the XC kernel itself. By relating approximate XC kernels to approximate 
local vertex corrections we then generate beyond-GW self-energies that include a 
discontinuity in the local vertex function. The quantitative importance of this result 
is illustrated with a numerical study of the local exchange vertex on model systems.
\end{abstract} 
\maketitle
\section{Introduction}
\label{intro}
Kohn-Sham (KS) density functional theory (DFT) is the most widely used 
many-electron approach for numerical simulations in physics, 
chemistry and materials science. \cite{kohnsham1965,rg84} 
In both its static and time-dependent (TD) version, the interacting many-electron 
problem is reformulated into the simpler problem of non-interacting electrons 
moving in a self-consistent, single-particle KS potential. This construction 
is formally exact, but is expected to become more constrained 
when the electrons are strongly correlated. On the other hand, 
exact studies on simple systems have shown that the KS potential 
captures effects of strong correlation by exhibiting rather 
intuitive peak and step features. \cite{stepsbarth,perdewsteps,gritsenko,helbigsteps,TDSTEPS,hodgson17} 
Standard local and semi-local approximations to the exchange-correlation (XC) 
energy, such as LDA and GGA functionals, miss these features and, as a consequence, 
tend to spread out or delocalize the electrons in the system.  

In a seminal work of Perdew et al.\cite{pplb82} it was shown that a key 
feature of the exact XC energy functional is 
a derivative discontinuity (DD) at integer particle numbers. 
The DD corrects the largely underestimated single-particle KS gap, 
and prevents delocalization through the formation 
of discontinuous XC steps in the KS potential. 
To capture strong correlation with (TD)DFT it is 
thus necessary to construct functionals with the DD. 
It should be noted that the concept of strong correlation can be ambiguous since it 
sometimes refers to effects beyond semi-local approximations in DFT and 
sometimes to multi-reference effects beyond Hartree-Fock (HF). 
In the first case the DD can often be captured by local exact-exchange,\cite{kli92,hg13} 
hybrid\cite{becke} or corrective DFT+U functionals.\cite{dftu} 
In the latter, truly correlated case, the DD constitute the entire gap, and it 
still remains a challenge to find approximate functionals.

Approximations based on many-body perturbation theory (MBPT) 
have shown to successfully capture many effects of correlation such as
screening, van der Waals forces and derivative discontinuities derived 
from nonlocal exchange. One example is the random phase approximation (RPA), 
derived from the GW self-energy within MBPT. The RPA and the GW 
approximation (GWA) share many virtues but both fail in describing localized states in 
strongly correlated systems such as Mott insulators. \cite{hrg12,GWRPA1,GWRPA,colonna2} 
The GWA can be improved by including so-called vertex corrections. 
So far a HF-like vertex has been employed to construct various total energy expressions 
based on partial re-summations of particle-hole diagrams. 
\cite{doi:10.1063/1.3250347,PhysRevB.88.035120,PhysRevB.92.081104,toulouse_prl,RPAx-F} 
Similarly, approximations based on the local exact-exchange vertex was studied in 
Refs. \onlinecite{hvb10,hesselmann_random_2010,PhysRevB.90.125150,gorlingrpax}. Local vertex 
corrections have also been studied at the simpler level of LDA. \cite{delsolevertex} 
Improved local vertex corrections can be derived from improved XC kernels within TDDFT. \cite{grosskohn,bruneval} 
The XC kernel is defined as the functional derivative of the 
XC potential with respect to the density and is the crucial quantity to calculate 
excitation energies within linear response TDDFT. \cite{pgg96} Recently, it was shown that also 
the XC kernel exhibits discontinuities at integer particle number, important 
for charge-transfer excitations\cite{hg12,hg13} and chemical reactivity indices. \cite{hgjcp12} 

In this work we will investigate the effects of the discontinuity of the XC kernel 
for constructing local vertex corrections to the many-body self-energy. It has, e.g., 
previously been shown that affinities need a three-point vertex to be accurately described. \cite{localvertexromaniello} 
Here we show that this problem can reformulated into the problem of a missing 
dynamical discontinuity of the two-point XC kernel. From the ACFD (adiabatic connection fluctuation dissipation) expression to the total 
energy we further derive an exact expression to the fundamental gap in terms of the XC kernel 
and its corresponding discontinuity. Finally, we calculate the discontinuity of the 
exact-exchange kernel and use it to determine correlated gaps beyond the GWA. 

\section{Derivative discontinuity in (TD)DFT}
In this section we review some of the mathematical formulas underlying 
the concept of the DD in static and time-dependent DFT. 

The DD refers to a discontinuous change in the derivative of the XC energy, $E_\xc$, 
as the density passes integer particle numbers. As a consequence, 
the corresponding XC potential, $v_\xc$, jumps with a space-independent constant. 
The DD has been proven to be a feature of the exact theory, 
\cite{pplb82,sagvolden,cohenmori09,SCEDD} and to play a key role when applying 
(TD)DFT to different physical problems 
such as, e.g., charge-transfer processes, Kondo or Coulomb blockade 
phenomena. \cite{TDSTEPS,hg12,CBDD,SFIDD,KONDO1,KONDO2,CTDD} The most direct
effect of the discontinuity is, however, seen when trying to extract the fundamental gap from DFT. 
The ionization energy $I$ and the affinity $A$ of a system with $N_0$ 
particles can be calculated as the left and right derivatives of 
the total energy with respect to particle number $N$, i.e, 
\be
-I=\left.\frac{\partial E}{\partial N}\right|_-,\,\,\,\,\,-A=\left.\frac{\partial E}{\partial N}\right|_+
\ee
where the subscript $\pm$ refers to the value of the quantity at $N^\pm_0=N_0+0^\pm$. 
The ionization energy of the KS system is equal to the ionization energy of the true 
interacting system \cite{almbarth} but the KS affinity $A_s$ has to be corrected 
with the discontinuity $\D_\xc$ of the XC potential. The gap of the interacting 
system is thus equal to the KS gap plus the discontinuity, i.e.,
\be
I-A=I-A_s+\D_\xc.
\label{gapia}
\ee
It has been demonstrated that the DD is comparable in size to 
the KS gap for both solids \cite{gunnshon,gruning06} and molecules. \cite{gapsfinite2,gapsfinite} Moreover, in strongly correlated 
Mott insulators the discontinuity accounts for the entire gap. \cite{yangscience,lorenzana} 

Following Ref.  \onlinecite{hg12} we will now derive a formal expression for the quantity $\D_\xc$ which is useful for extracting 
the discontinuity of approximate functionals. Using the chain rule we can write 
the partial derivative of $E_\xc$ with respect to $N$ as
\bea
\frac{\partial E_{\xc}}{\partial N}=\int\! d \vvr \,  v_{\xc}(\vvr )\frac{\partial n(\vvr )}{\partial N}.
\label{deriv1}
\eea 
In the limit $N_0^-=N_0+0^-$ this expression can be rewritten as
\bea
0=\left.\frac{\partial E_{\xc}}{\partial N}\right|_--\int \!d \vvr \, v^-_{\xc}(\vvr )f^-(\vvr ),
\label{fix-}
\eea
where $f(\vvr )=\partial n(\vvr )/\partial N$ is the so-called Fukui function. \cite{fukui1,Fukui} Let us now look
at the right derivative of $E_\xc$, which will be different from the left derivative if a 
DD is present. From Eq. (\ref{deriv1}) we see that this difference can appear either in 
the XC potential $v^+_{\xc}(\vvr )=v^-_{\xc}(\vvr )+\D_\xc$ or in the Fukui function. 
In fact, it is easy 
to see that in the case of local or semilocal functionals such as LDA and PBE the discontinuous 
behaviour is restricted to the Fukui function, i.e., $\D_\xc=0$. 
Meta-GGAs have shown to exhibit a discontinuity in the XC potential, albeit very small. \cite{eichhellgren} 
On the other hand, orbital functionals based on MBPT are known to accurately capture the discontinuity 
of the XC potential.
 
Let us now write $v_\xc(\vvr )=v^-_\xc(\vvr )+\D_\xc(\vvr )$ and cast Eq. (\ref{deriv1}) into  
\bea
\int\! d\vvr \, \D_\xc(\vvr )f(\vvr )=\frac{\partial E_{\xc}}{\partial N}-\int \!d \vvr \, v^-_{\xc}(\vvr )f(\vvr ).
\label{deriv2}
\eea
In the limit $N\rightarrow N_{0}^+$, $\D_\xc(\vvr )\to\D_\xc$ and we find an expression for the 
discontinuity of $v_\xc$
\bea
\D_\xc=\left.\frac{\partial E_{\xc}}{\partial N}\right|_+-\int \!d \vvr \, v^-_{\xc}(\vvr )f^+(\vvr ).
\label{disc}
\eea

A systematic scheme for generating orbital functionals based on MBPT was presented in 
Ref. \onlinecite{vbdvls05}. The idea is to start from variational energy expressions of the 
Green's function $G$ and then restrict the variational freedom to Green's functions $G_s$ 
coming from a local KS potential. One can, for example, start from the 
Klein functional \cite{klein}
\bea
E[G]&=&-i\F[G]+E_{\rm H}+i\Tr [GG_s^{-1}-1+\ln (-G^{-1})],
\label{klein}
\eea
which contains another functional $\F$ having the property that the self-energy $\S$ is 
generated as $\S=\d\F/\d G$. When restricting to KS Green's functions it is easy to see 
that the XC energy corresponds to $E_\xc=-i\F[G_s]$. Furthermore, the equation for the XC potential is 
nothing but the linearized Sham-Schl\"uter (LSS) equation \cite{sham,shams,godbylss,godbylss2,casida95}
\be
\int \!d 2 \,\chi_s(1,2) v_\xc(2)=\int \!d2d3\,\S_{\rm xc}(2,3) \L(3,2;1).
\label{lsseq}
\ee
Here we have simplified the notation by introducing $1=\vvr _1,t_1$.
Even with the exact self-energy this scheme will never be exact but it 
has shown to produce useful approximations such as the exact-exchange (EXX) approximation 
(based on the HF self-energy)  and the random phase approximation (RPA) (based on the GW self-energy). 
To determine the discontinuities of the Klein XC potentials, Eq. (\ref{klein}) and Eq. (\ref{lsseq}) 
must be generalized to ensemble Green's functions. This was done in Refs. \onlinecite{hg12,hrg12} showing that 
the discontinuities of the Klein functional are given by
\bea
\D_\xc&=&\int \!d\vvr  d\vvr ' \,\vf^*_{\rm L}(\vvr )\S^+_{\xc}(\vvr ,\vvr ',\ve^+_{\rm L})\vf_{\rm L}(\vvr ')\nn\\
&&\,\,\,-\int \!d \vvr \, v^-_{\xc}(\vvr )|\vf_{\rm L}(\vvr )|^2.
\label{discklein}
\eea
where $\rm L$ signifies the LUMO (lowest unoccupied molecular orbital). 
The superscript on the self-energy can be dropped as it is, in general, invariant with respect to 
a constant shift of the potential. Combining Eq. (\ref{discklein}) with Eq. (\ref{gapia}) 
leads to 
\be
-A=\varepsilon^{-}_{\rm L}+\D_\xc=\varepsilon^{-}_{\rm L}+\bra \vf_{\rm L}|\S_{\rm xc}(\varepsilon_{\rm L})-v^-_{\rm xc}| \vf_{\rm L}\ket
\label{affdisc}
\ee
where the KS LUMO is calculated from $v^-_{\rm xc}$. This equation is very similar to the 
first order quasi-particle equation within MBPT.

The above mathematical analysis was restricted to the static case but also the time-dependent 
$v_\xc$ exhibits jumps. In the linear response regime this leads to a frequency 
and space-dependent discontinuity in the XC kernel given by
\be
f^+_\xc(\vvr ,\vvr ',\w)=f^-_\xc(\vvr ,\vvr ',\w)+g_\xc(\vvr ,\w)+g_\xc(\vvr ',\w).
\ee
In order to determine the discontinuity $g_\xc(\vvr ,\w)$ of the XC kernel, given an approximation to $v_\xc$, 
we use a similar procedure as in the static case but with a generalized ensemble that allow densities 
to change particle number in time. In Ref. \onlinecite{hg12} it was shown that such an ensemble was necessary 
for functional derivatives to be uniquely defined. 

Assuming that $v_\xc$ is defined on such a generalised domain of densities we can evaluate 
the functional derivative of $v_\xc$ with respect to the time-dependent number of particles, i.e.,  
\bea
\left.\frac{\d v_\xc(\vvr  t)}{\d N(t')}\right |_{n^+_0}&=&\int \!d\vvr '\, f^-_\xc(\vvr ,\vvr ', t-t')f^+(\vvr ')\nonumber\\
&&\!\!\!\!\!\!\!\!\!\!\!\!\!\!\!\!\!\!\!\!\!\!+\int \!d\vvr ' \,g_\xc(\vvr ', t-t')f^+(\vvr ')+g_\xc(\vvr , t-t').
\label{disctime}
\eea
Let us now specialize to functionals derived from the Klein action functional of MBPT. 
The TD XC potential $v_\xc$ is then obtained from the LSS equation (Eq. (\ref{lsseq}) above). 
Combined with Eq. (\ref{disctime}), we find 
the following equation to determine $g_\xc$ 
\bea
\int \!d 2 \,\chi_s(1,2) g_\xc(2,t)&=&\nonumber\\
&&\!\!\!\!\!\!\!\!\!\!\!\!\!\!\!\!\!\!\!\!\!\!\!\!\!\!\!\!\!\!\!\!\!\!\!\!\!\!\!\!\!\!\!\!\!\!\!\!\!\!\!\!\!\!\!\left.\int \!d2d3\,\S_{\rm xc}(2,3)\frac{\d \L(3,2;1)}{\d N(t)}\right|_{n^+_0}\nonumber\\
&&\!\!\!\!\!\!\!\!\!\!\!\!\!\!\!\!\!\!\!\!\!\!\!\!\!\!\!\!\!\!\!\!\!\!\!\!\!\!\!\!\!\!\!\!\!\!\!\!\!\!\!\!\!\!\!+\left.\int \!d(2345)\,\frac{\d \S_{\rm xc}(2,3)}{\d G_s(4,5)}\frac{\d G_s(4,5)}{\d N(t)}\right|_{n^+_0}\L(3,2;1)\nn\\
&&\!\!\!\!\!\!\!\!\!\!\!\!\!\!\!\!\!\!\!\!\!\!\!\!\!\!\!\!\!\!\!\!\!\!\!\!\!\!\!\!\!\!\!\!\!\!\!\!\!\!\!\!\!\!\!-\!\int\! d2d3\,\chi_s(1,2) f^-_{\xc}(2,3)\left.\frac{\d n(3)}{\d N(t)}\right|_{n^+_0}\nn\\
&&\!\!\!\!\!\!\!\!\!\!\!\!\!\!\!\!\!\!\!\!\!\!\!\!\!\!\!\!\!\!\!\!\!\!\!\!\!\!\!\!\!\!\!\!\!\!\!\!\!\!\!\!\!\!\!-\!\left.\int \!d 2 \,
v^+_\xc(2)\frac{\d\chi_s(1,2)}{\d N(t)}\right|_{n^+_0}.
\label{discfxc}
\eea

The discontinuity of the XC kernel was in Ref. \onlinecite{hg12} shown to be finite and carry a strong
frequency dependence. Already the static discontinuity turned out to give a significant contribution 
when calculating reactivity indices such as the Fukui function. \cite{hgjcp12} In the next 
sections we will show that the dynamical discontinuity give a significant contribution when calculating beyond-GW 
gaps from local vertex corrections within MBPT.
\section{Local vertex corrections from $f_\xc$}
The most popular approximation to the self-energy is the GWA, 
in which the bare Coulomb interaction $v$ of the 
HF approximation is replaced by a dynamically screened Coulomb interaction, $W$. \cite{hedin,ferdirevgw,luciarevgw} 
Any effect beyond the GWA is usually referred to as a vertex correction, denoted by $\G$, 
and the exact self-energy can be written as 
\be
\S=iGW\G
\label{selfv}
\ee 
where 
\be
W=v+vPW,\,\,\,\,\,P = -iGG\G\,\,
\label{wphedin}
\ee 
and $P$ is the so-called irreducible polarisability. The vertex function $\G$ is defined as 
\be
\G=-\frac{\d G^{-1}}{\d V}=1+\frac{\d \S}{\d V}
\label{defvert}
\ee
where $V=v_{\rm ext} + v_{\rm H}$ is the sum of the external and Hartree potential. 
If the vertex is set to 1 we obviously obtain the GWA. From the definition of the vertex 
we see that the next level of approximation can be generated iteratively from the 
derivative of the GW self-energy. This self-energy is expected to be very accurate but too 
complex to be applied to a real systems. 

With the idea of approximating the numerically cumbersome three-point vertex function it has been 
suggested to replace the self-energy in Eq. (\ref{defvert}) by local XC potentials derived from TDDFT. 
This generates local vertex functions depending only on two space and time variables. \cite{delsolevertex,bruneval} 
The LDA potential was used in Ref. \onlinecite{delsolevertex} 
but produced only a very small change to the GW gaps. Moreover, in Ref. \onlinecite{localvertexromaniello} 
it was found that whereas ionization energy were well captured by local vertex corrections 
affinities were not. In order to overcome these limitations we will now look at more advanced 
XC potentials derived from the Klein MBPT scheme in the previous section, a scheme which can also 
be used to derive local vertex functions. These approximations are, e.g., guaranteed to be conserving due 
to the $\F$-derivability of the allowed self-energies, \cite{baym2,baym1} a potentially important feature.

In general, we can write any local vertex function as
\be
{\rm\G}_{\rm xc}= 1 + \frac{\d v_{\rm xc}}{\d V}=\frac{1}{1-{\rm\G}_\xc^1}
\label{defvertloc}
\ee
where
\be
{\rm\G}_\xc^1=\frac{\d v_\xc}{\d V_s},\,\,\,\,V_s=v_{\rm ext}+v_{\rm Hxc}.
\label{vert1}
\ee
Using the chain rule, the local vertex function is easily related to the XC kernel of TDDFT
\be
{\rm\G}_\xc^1 =\frac{\d v_\xc}{\d n}\frac{\d n}{\d V_s}=f_\xc\chi_s.
\label{vert1fxc}
\ee
We can now insert the local vertex of Eq. (\ref{defvertloc}) into Eq. (\ref{selfv}) in order to 
generate an approximate beyond-GW self-energy
\be
\S=iGW\G_{\rm xc}.
\label{selfvloc}
\ee 
There are now two important issues to consider. Firstly, when calculating the affinity 
the self-energy should be evaluated at $N_0^+$. This is similar to the case of approximating 
the nonlocal self energy with a local XC potential, where the latter is discontinuous at $N_0$. 
Since the ensemble XC kernel also has discontinuities, from Eq. (\ref{vert1fxc}) we see that the 
local ensemble vertex function must have discontinuities 
\be
\G_{\rm xc}^+=\G_{\rm xc}^-+G_{\rm xc}.
\label{discvert}
\ee
For example, looking at the first order term in the expansion of the vertex correction 
(Eq. (\ref{vert1fxc})) we see immediately that 
\be
G^1_{\rm xc}=g_\xc\chi_s
\label{gxc}
\ee
with $g_\xc$ as defined in Eq. (\ref{discfxc}) above. 

Secondly, we notice that the self-energies used in Eq. (\ref{discklein}) should be $\F$-derivable whereas 
self-energies from Eq. (\ref{selfvloc}) are, in general, not. We can, however, derive a very 
similar expression to Eq. (\ref{discklein}) starting from the exact ACFD-formula to the 
total energy 
\be
E=E_s +\frac{i}{2}\int \frac{d\w}{2\pi}\int d\lambda{\rm Tr}\{v[\chi_{\lambda}(\omega)-\delta n]\}.
\label{echi}
\ee
Here, $E_s$ is the total energy in the Hartree approximation and $\chi_{\lambda}$ is the scaled density 
correlation function.  We also define $\Tr\{ AB \}=\int d {\bf r} d {\bf r}' A({\bf r},{\bf r}')B({\bf r}',{\bf r})$ and 
$\delta n=\d({\bf r},{\bf r}')n({\bf r})$.
With the definition of the exact local vertex function in Eq. (\ref{defvertloc}) it is easy to rewrite 
Eq. (\ref{echi}) in terms of Eq. (\ref{selfvloc}). \cite{niquet} 
We find
\be
E=E_s-\frac{i}{2}\int \frac{d\w}{2\pi}\int \frac{d\lambda}{\lambda}{\rm Tr}\{\S^{GW\G_{\xc}}_{\lambda}G_s\}.
\label{eselflocal}
\ee
We see that for the total energy the discontinuity of the local vertex function will always vanish. 
Let us now take the derivative of the XC part with respect to particle number. We find
\bea
\left.\frac{\partial E_{\rm xc}}{\partial N}\right|_+&=&-i\int_0^1\! \frac{d\lambda}{\lambda}\,\Tr\! \left\{\left.\!\tilde\S^{GW\G_{\xc}^+}_{\lambda}\frac{\partial G_s}{\partial N}\right|_+\right.\nn\\
&&\,\,\,\,\,\,\,\,\,\,\,\,\,\,\,\,\,\,\,\,\,\,\,\,\,\,\,\,\,\,\,\,\,\,\,\,\,\,\,\,\,\,-\left.\left.\frac{\l}{2}\chi^\l v\chi^\l \frac{\partial f^\l_{\rm xc}}{\partial N}\right|_+\!\right\}.
\label{acfdtdisc}
\eea
This expression together with Eq. (\ref{disc}) yields an exact expression for 
the discontinuity of the XC potential and hence the fundamental gap.
We see that the first term is very similar to Eq. (\ref{discklein}) but with 
the exact nonlocal self-energy replaced by a symmetrized self-energy with a local vertex function
\be
\tilde\S^{GW\G_{\xc}^+}=iG_s\G^+_\xc Wv^{-1}W\G^+_\xc.
\label{symself}
\ee
Within the RPA ($\G_\xc=1$) the $\l$-integral can be carried out analytically 
and this term becomes exactly Eq. (\ref{discklein}) with the GW self-energy. 
It is clear that if $\G_\xc$ is discontinuous the self-energy in Eq. (\ref{symself}) 
will be discontinuous, with important consequences for calculating 
fundamental gaps. In the next section we will quantify its contribution.
\section{Correlated gaps from the EXX vertex}
In this section we will study the EXX approximation to the XC kernel, 
which can be considered the most simple, yet consistent, vertex correction for going beyond the GWA. 
The EXX approximation has been shown to support a discontinuity in both the 
EXX potential \cite{kliexx} and the EXX kernel, \cite{hg12} necessary for EXX density to equal the HF density 
to first order, at equilibrium and in linear response, respectively. 

The fully nonlocal HF vertex is defined as
\be
\G_{\rm HF}= 1 + \frac{\d \S_{\rm HF}}{\d V},
\label{defhfvert}
\ee
a three point function in time and space. The EXX local vertex 
function can instead be written as
\be
\G_{\rm x}= 1 + \frac{\d v_{\rm x}}{\d V}=\frac{1}{1-\G_\x^1},
\label{defexxvert}
\ee
where $v_\x$ is the local time-dependent EXX potential given by 
the LSS equation (Eq. (\ref{lsseq})) within the HF approximation to the self-energy. 
The first order term in the expansion of the EXX vertex is given by
\be
\G_\x^1=\frac{\d v_\x}{\d V_s},\,\,\,\,V_s=v_{\rm ext}+v_{\rm Hx}.
\label{vert1}
\ee
Using the chain rule we can write the EXX kernel ($f_\x$) as
\be
f_\x=\frac{\d v_\x}{\d n}=\G_\x^1 \chi_s^{-1}.
\ee
This kernel has been studied in several previous works. It is known to produce excellent total energies when used 
in the ACFD total energy expression of Eq. (\ref{echi}). \cite{hvb10,hesselmann_random_2010} It captures 
excitonic effects \cite{kgexcitons} and charge-transfer excitations within the single-pole 
approximation. \cite{hg12} 

In this work we have calculated the EXX kernel and the corresponding EXX vertex of 
one-dimensional soft-Coulomb systems using a precise and efficient 
spline-basis set. \cite{bachau,hvb07,hg13} Below we will study the effect the exchange vertex when 
calculating the fundamental gap of molecular systems. We will also compare 
the fully nonlocal approximation (Eq. (\ref{defhfvert})) to 
the local approximation (Eq. (\ref{defexxvert})). 
\subsection{MP2}
In order to test the theory we start by looking at the most 
simple approximation to the self-energy that can be written in terms of the 
exchange vertex. Expanding the self-energy to second order in the Coulomb interaction 
we obtain what is known as the the second order Born or MP (M\o ller-Plesset) approximation, \cite{jemp2} 
in which the sum of first and second order exchange can be written in terms of the first order HF vertex function 
($\G_{\rm HF}^1$), i.e., 
\be
\S^{\rm MP2}=i\G_{\rm HF}^1vG_s+iv\chi_svG_s.
\label{mp2}
\ee 
The last term is a second order term related to the screened 
interaction. In the appendix we evaluate these terms explicitly (Eq. (\ref{term1}) and 
Eq. (\ref{term2})), and we see that they are smooth as functions of particle number $N$. 
They are also invariant with respect to a constant shift of the external potential.

We will now approximate this self-energy with the local EXX vertex
defined above. We thus write 
\be
\S_{\rm L}^{\rm MP2}=i\G_{\x}^1vG_s+iv\chi_svG_s.
\ee
Since the EXX vertex is discontinuous due to a discontinuity of the EXX kernel
we have to evaluate the first term ($\S_{\rm \G_{\rm x}^1}$) 
differently depending on if we are calculating affinities or 
ionization energies. In the limit $N=N_0^-$, relevant for ionization energies, 
we write 
\be
\S^-_{\rm \G_{\rm x}^1}=if_\x^-\chi_svG_s,
\label{term-}
\ee
and, in the limit $N=N_0^+$, relevant for affinities, we write
\be
\S^+_{\rm \G_{\rm x}^1}=if_\x^-\chi_svG_s+ig_\x\chi_svG_s.
\label{term+}
\ee
The discontinuity $g_\x$ gives rise to a second non-vanishing term. 
Furthermore, if we take the Fourier transform of Eq. (\ref{term+}) we see that the frequency 
dependence in $g_\x$ has to be taken into account
\bea
\S^+_{\rm \G_{\rm x}^1}(\vvr ,\vvr ',\w)&=&i\int\! \frac{d\w'}{2\pi}\,G_s(\vvr ,\vvr ',\w-\w')\nn\\
&&\!\!\!\!\!\!\!\!\!\!\!\!\!\!\!\!\!\!\!\!\!\!\!\!\!\!\!\!\!\times \left[\int \!d\vvr _1d\vvr _2\,v(\vvr ,\vvr _2)\chi_s(\vvr _2,\vvr _1,\w')f^-_\x(\vvr _1,\vvr ',\w')\right.\nn\\
&&\!\!\!\!\!\!\!\!\!\!\!\!\!\!\!\!\!\!\!\!\!\!\!\!\!\!\!\!\!\left.+ \int \!d\vvr _1d\vvr _2\,v(\vvr ,\vvr _2)\chi_s(\vvr _2,\vvr _1,\w')g_\x(\vvr _1,\w')\right].
\label{fourmp2}
\eea

Let us now specialize to a two-electron system. The EXX kernel at $N=N_0^-$ is then simply 
given by 
\be
f^-_\x(\vvr _1,\vvr ',\w')=-\frac{1}{2}v(\vvr _1,\vvr ').
\ee
If we insert this kernel into the first term of Eq. (\ref{fourmp2}) we see that 
it is equal to minus one half of the second term of Eq. (\ref{mp2}).
The second term of Eq. (\ref{fourmp2}), the term containing the discontinuity, 
can be evaluated with the help of Eq. (\ref{discfxc}), adapted to EXX. 
In the appendix we have explicitly evaluated this term for a two-electron system (see Eq. (\ref{term3})).
\begin{figure}
\resizebox{0.98\columnwidth}{!}{\includegraphics{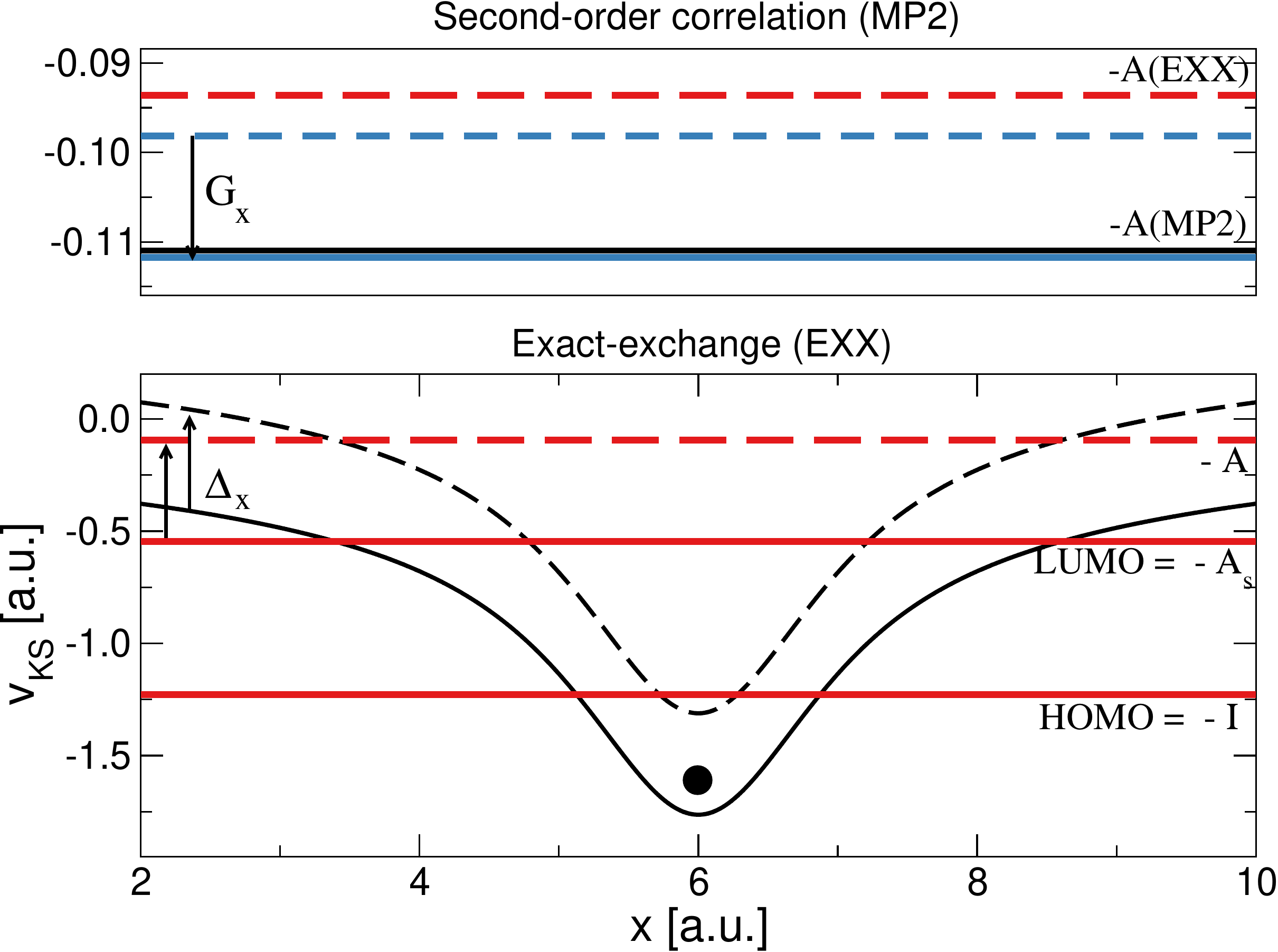}}
\caption{Lower panel: KS potential of a soft-Coulomb two-electron atom in the EXX approximation 
(black full and dashed curves). Horizontal lines indicates the KS HOMO and LUMO energy levels (red full)
and the corrected affinity (red dashed). Upper panel: MP2 corrected affinity (blue dashed and full lines 
are without and with discontinuity, respectively). Black full line is the nonlocal MP2 result. }
\label{fig:1}       
\end{figure}
\begin{figure}
\resizebox{0.98\columnwidth}{!}{\includegraphics{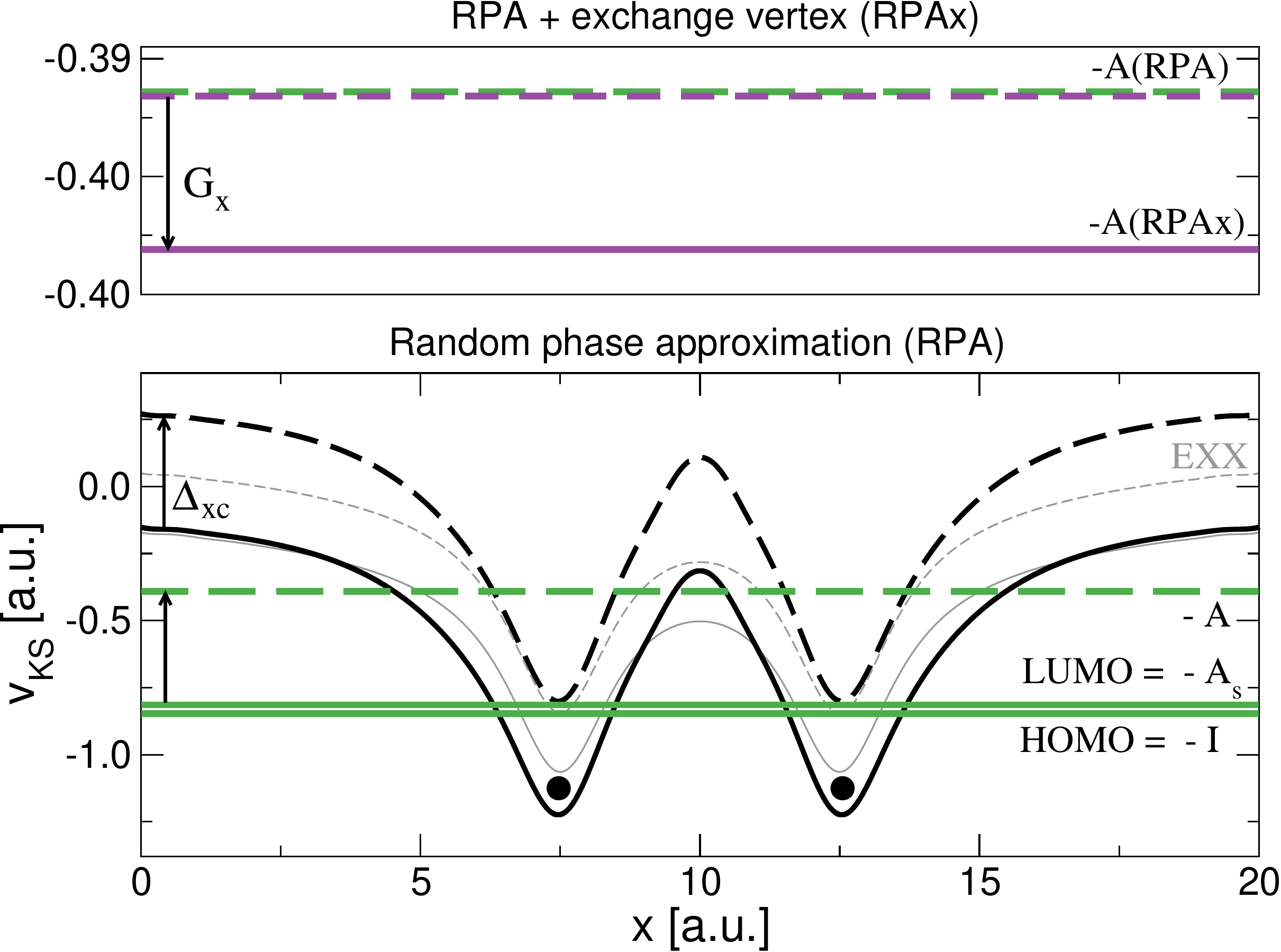}}
\caption{Lower panel: KS potential of a soft-Coulomb stretched H$_2$ molecule in the RPA 
approximation (black full and dashed curves). Horizontal lines indicates the KS HOMO and 
LUMO energy levels (green full) and the corrected affinity (green dashed). EXX approximation in the background (grey curves). 
Upper panel: RPAx corrected affinity (purple dashed and full lines are without and with discontinuity, respectively).}
\label{fig:2}       
\end{figure}

In Fig.~\ref{fig:1} we present the results for a soft-Coulomb atomic system 
containing two electrons. The black dot signifies the location of the atom. 
In the lower panel we plot the EXX KS potential (full black line) 
obtained by imposing Eq. (\ref{fix-}), which ensures the correct asymptotic limit of the 
potential. The dashed black line is the same potential but shifted by the EXX discontinuity 
$\D_\x$ (Eq. (\ref{discklein})). This is the result one would get by evaluating the potential 
at $N=N_0^+$, as indicated to left with the arrows pointing upwards. The location of the  
KS HOMO (highest occupied molecular orbital) ($-I$) and the KS LUMO (lowest unoccupied molecular orbital) 
($-A_s$) are indicated with full horizontal red lines. The red dashed horizontal line 
indicates the true affinity ($-A$) after correction with the discontinuity (Eq. (\ref{affdisc})), also indicated 
by an arrow pointing upwards to the left. 

The upper panel shows the correction to the affinity due to MP2 correlation. The red dashed A(EXX) line 
is the EXX affinity (a duplicate of the $-A$ line in the lower panel) and the blue dashed line 
is the correction due to the first term of Eq. (\ref{term+}). Keeping only this term corresponds to using 
Eq. (\ref{term-}) to calculate affinities. Including the discontinuity 
term (second term in Eq. (\ref{term+}), here called $G_\x$) further shifts the affinity (blue full 
horizontal line). Actually, we find that 
$G_\x$ corresponds to 75\% of the total correction. The black full line in the same panel denoted A(MP2), 
is obtained by using the fully nonlocal vertex, i.e., by evaluating Eq. (\ref{mp2}). We see that the black 
and blue full lines almost coincide. We can thus conclude that in order to reproduce the nonlocal MP2 
approximation the discontinuity (i.e. the $G_\x$ term) of the EXX kernel is crucial.
\subsection{RPAx}
We will now construct a more advanced self-energy that takes into account both screening and vertex 
corrections to all orders in the Coulomb interaction. In this way, we will consistently 
incorporate the EXX vertex in both the screened interaction (Eq. (\ref{wphedin})) and in the 
self-energy (Eq. (\ref{selfv})). We call this self-energy RPAx
\be
\S^{\rm RPAx}=iG_sW_\x\G_{\x}.
\ee 
This is the self-energy (although symmetrized) one obtains from the ACFD-formula 
including the EXX kernel (i.e. the RPAx energy) but by ignoring variations of $f_\x$ with 
respect to the density. \cite{hvb10} It can also be seen as the local approximation to the 
self-energy that includes vertex corrections at the time-dependent HF level. \cite{toulouse_prl} 

We applied this approximation to the stretched hydrogen molecule, for which we expect screening
to be more important. In the lower panel of Fig.~\ref{fig:2} we plot the KS potential at the 
RPA level (black full line). We also show the EXX potential in the background (grey thin line). 
The KS HOMO-LUMO gap is vanishing small but if we add the RPA discontinuity $\D_\xc$ (obtained from 
Eq. (\ref{discklein}) with the GW self-energy) the affinity shifts substantially. We also note that it is larger than the corresponding EXX correction, 
in agreement with the analysis in Ref. \onlinecite{GWRPA}. Neither the HF nor the GWA is able 
to correctly describe the gap of stretched H$_2$ since it is a strongly correlated 
Mott-like system. \cite{GWRPA1} Including exchange effects in the vertex is not expected 
to qualitatively improve upon the GWA. Indeed, in the upper panel we present the results from the RPAx self-energy
and we see that the correction is of the wrong sign. However, in this case, we see that the 
discontinuity is even more important to reproduce 
the nonlocality of the self-energy, accounting for almost the entire correction. 

\section{Conclusions}
In this work we have derived approximate self-energies based on local vertex corrections 
derived from XC kernels within the Klein MBPT formulation of TDDFT. These vertex 
corrections capture a dynamical discontinuity of the XC kernel, needed to 
accurately describe electron affinities. A numerical study on model molecular 
systems shows that the discontinuity of the local vertex corresponds to the 
largest correction to the GW affinity. 

From the ACFD formula to the total energy we further show that an exact expression 
for the discontinuity of the XC potential can be written in terms of the XC kernel 
and its derivative. Although this work focused on the DD as a correction 
to the gap, a very important consequence of the discontinuity is to localise electrons 
in strongly correlated systems. The ability of an XC potential derived from the ACFD 
expression to do so will strongly rely on the discontinuous nature of the XC kernel.  
\section{Appendix}
Here we give explicit expressions for the formulas in Sec.~4. The MP2 self-energy is composed of two terms.
The second term in Eq.~(\ref{mp2}) is a first order screening correction so we denote it with a 
subscript $W_1$. In terms of orbitals and eigenvalues it is given by
\bea
\S_{\rm W_1}(\vvr ,\vvr ',\w)&=&2\sum_{kq}\varphi_k(\vvr )\int d\vvr _1 v(\vvr ,\vvr _1)f^*_q(\vvr _1)\nn\\
&&\!\!\!\!\!\!\!\!\!\!\!\!\!\!\!\!\!\!\!\!\!\!\!\!\!\!\!\times \varphi^*_k(\vvr ')\int d\vvr _1 v(\vvr ',\vvr _1)f_q(\vvr _1)\nn\\
&&\!\!\!\!\!\!\!\!\!\!\!\!\!\!\!\!\!\!\!\!\!\!\!\!\!\!\!\times \left[\frac{n_k}{\w-\ve_k+Z_q-i\eta}+\frac{1-n_k}{\w-\ve_k-Z_q+i\eta}\right],
\label{term1}
\eea
where $q$ is the particle-hole index, $f_q$ is the 'bare' excitation function given by a product of occupied 
and unoccupied KS orbitals, and $Z_q$ is the corresponding eigenvalue difference. 
The first 'vertex' term is instead given by
\bea
\S_{\rm \G_{\rm HF}^1}(\vvr ,\vvr ',\w)&=&\sum_{ksp}\varphi^*_k(\vvr )\int d\vvr _1 v(\vvr ,\vvr _1)\varphi_s(\vvr _1)\varphi_p(\vvr _1)\nn\\
&&\!\!\!\!\!\!\!\!\!\!\!\!\!\!\!\!\!\!\!\!\!\!\!\!\!\!\!\!\!\!\!\!\!\!\!\!\!\!\!\!\!\!\!\times \,\,\varphi_s(\vvr ')\int d\vvr _1 v(\vvr ',\vvr _1)\varphi_k(\vvr _1)\varphi^*_p(\vvr _1)\times(-1)\nn\\
&&\!\!\!\!\!\!\!\!\!\!\!\!\!\!\!\!\!\!\!\!\!\!\!\!\!\!\!\!\!\!\!\!\!\!\!\!\!\!\!\!\!\!\!\times \left[\frac{n_kn_s(1-n_p)}{\w-\ve_s+\ve_p-\ve_k-i\eta}+\frac{(1-n_k)n_p(1-n_s)}{\w-\ve_s+\ve_p-\ve_k+i\eta}\right].
\label{term2}
\eea
The sum of the terms (Eq. (\ref{term1}) and Eq. (\ref{term2})) constitute the MP2 self-energy. The local MP2 self-energy also contains
the term in Eq. (\ref{term1}) and in addition two local terms, one containing $f_\x$ and one its discontinuity $g_\x$. For a 
two-electron system the $f_\x$-term is just minus one half of Eq. (\ref{term1}). The second term is generated from Eq. (\ref{discfxc}) yielding
\bea
ig_\x\chi_svG_s&=&\sum_{k,p\ne L}\varphi_k(\vvr ')\int \! d\vvr _1 \,v_\x(\vvr _1)\varphi_p(\vvr _1)\varphi^*_L(\vvr _1)\nn\\
&&\!\!\!\!\!\!\!\!\!\!\!\!\!\!\!\!\!\!\!\!\!\!\!\!\!\!\!\times \varphi^*_k(\vvr )\int \!d\vvr _1 \,v(\vvr ,\vvr _1)\varphi_L(\vvr _1)\varphi^*_p(\vvr _1)\nn\\
&&\!\!\!\!\!\!\!\!\!\!\!\!\!\!\!\!\!\!\!\!\!\!\!\!\!\!\!\times \left[\frac{n_k}{\w+\varepsilon_L-\ve_p-\varepsilon_k-i\eta}+\frac{1-n_k}{\w-\ve_L+\ve_p-\ve_k+i\eta}\right]\nn\\
&&\!\!\!\!\!\!\!\!\!\!\!\!\!\!\!\!\!\!\!\!\!\!\!\!\!\!\!+\sum_{s,k,p\ne L}\varphi_k(\vvr ')n_s\int \! d\vvr _1 d\vvr _2\varphi_s(\vvr _1)\varphi^*_s(\vvr _2)\,v(\vvr _1,\vvr _2)\varphi_p(\vvr _1)\varphi^*_L(\vvr _2)\nn\\
&&\!\!\!\!\!\!\!\!\!\!\!\!\!\!\!\!\!\!\!\!\!\!\!\!\!\!\!\times \varphi^*_k(\vvr )\int \!d\vvr _1 \,v(\vvr ,\vvr _1)\varphi_L(\vvr _1)\varphi^*_p(\vvr _1)\nn\\
&&\!\!\!\!\!\!\!\!\!\!\!\!\!\!\!\!\!\!\!\!\!\!\!\!\!\!\!\times \left[\frac{n_k}{\w+\varepsilon_L-\ve_p-\varepsilon_k-i\eta}+\frac{1-n_k}{\w-\ve_L+\ve_p-\ve_k+i\eta}\right]\nn\\
&&\!\!\!\!\!\!\!\!\!\!\!\!\!\!\!\!\!\!\!\!\!\!\!\!\!\!\!+\sum_{kq}\varphi_k(\vvr )\int d\vvr _1 v(\vvr ,\vvr _1)f^*_q(\vvr _1)\nn\\
&&\!\!\!\!\!\!\!\!\!\!\!\!\!\!\!\!\!\!\!\!\!\!\!\!\!\!\!\times \varphi^*_k(\vvr ')\int d\vvr _1d\vvr _2 |\vf_L(\vvr _2)|^2v(\vvr _2,\vvr _1)f_q(\vvr _1)\nn\\
&&\!\!\!\!\!\!\!\!\!\!\!\!\!\!\!\!\!\!\!\!\!\!\!\!\!\!\!\times \left[\frac{n_k}{\w-\ve_k+Z_q-i\eta}+\frac{1-n_k}{\w-\ve_k-Z_q+i\eta}\right]\nn\\
&&\!\!\!\!\!\!\!\!\!\!\!\!\!\!\!\!\!\!\!\!\!\!\!\!\!\!\!+\sum_{kq}\varphi_k(\vvr )\int d\vvr _1 v(\vvr ,\vvr _1)\vf_s(\vvr _1)\vf^*_p(\vvr _1)\nn\\
&&\!\!\!\!\!\!\!\!\!\!\!\!\!\!\!\!\!\!\!\!\!\!\!\!\!\!\!\times \varphi^*_k(\vvr ')\int d\vvr _1d\vvr _2 \vf^*_L(\vvr _1)\vf_L(\vvr _2)v(\vvr _2,\vvr _1)\vf^*_s(\vvr _1)\vf_p(\vvr _2)\nn\\
&&\!\!\!\!\!\!\!\!\!\!\!\!\!\!\!\!\!\!\!\!\!\!\!\!\!\!\!\times \left[\frac{n_kn_s(1-n_p)}{\w-\ve_k+\ve_p-\ve_s-i\eta}+\frac{(1-n_k)n_s(1-n_p)}{\w-\ve_k-\ve_p-\ve_s+i\eta}\right]
\label{term3}
\eea

 \bibliographystyle{epj}

%
%
%

\end{document}